\documentstyle[12pt]{article}
\newcommand{\be}{\begin{eqnarray}}
\newcommand{\ee}{\end{eqnarray}}
\newcommand{\ba}{\begin{array}}
\newcommand{\ea}{\end{array}}
\newcommand{\nn}{\nonumber}
\newcommand{\ra}{\rightarrow}

\def\o{\over}
\def\l{\lambda}
\def\b{\beta}
\def\bl{\b_\l}
\def\bm{\b_M}
\def\f{\varphi}
\def\G{\Gamma}
\def\gf{\gamma_\f}
\def\fb{\bar\f}
\def\fm{\bar\f_0}
\def\p{\partial}
\def\Zf{Z_\f}
\def\loop#1{\hbox{$\raisebox{0.8pt}{$\bigcirc$}\kern-9pt{#1}\kern+5pt$}}

\begin{document}
\begin{flushright}
DIAS-STP-94-33\\
ITFA-94-29\\
ICN-UNAM-94-07\\
\today
\end{flushright}
\vskip 22 truept
\begin{center}
{\LARGE
Critical Temperature and Amplitude Ratios from a
Finite-Temperature Renormalization Group}
\vskip\baselineskip
{\bf M.A.\ van Eijck}\footnote{Institute for Theoretical Physics,
University of Amsterdam,
Valckenierstraat 65, NL-1018 XE Amsterdam, Netherlands},
{\bf Denjoe\ O' Connor}\footnote{School of Theoretical Physics,
Dublin Institute for Advanced Studies, 10 Burlington Road, Dublin 4,
Ireland.},
{\bf and C.R.\ Stephens}${}^{2,}$\footnote{Instituto de Ciencias Nucleares,
U.N.A.M., A. Postal 70-543, 04510 Mexico D.F., Mexico.}.
\end{center}
\vskip 1.3truein
{\bf Abstract:}
We study $\l\f^4$ theory using an environmentally friendly
finite-temperature renormalization group.
We derive flow equations, using a fiducial temperature as flow parameter,
develop them perturbatively in an expansion free from ultraviolet and
infrared divergences, then integrate them numerically from zero
to temperatures above the critical temperature.
The critical temperature, at which the mass vanishes,
is obtained by integrating the flow equations and is determined
as a function of the zero-temperature mass and coupling.
We calculate the field expectation value and minimum of the effective
potential as functions of temperature and derive some universal amplitude
ratios which connect the broken and symmetric phases of the theory.
The latter are found to be in good agreement with those of the
three-dimensional Ising model obtained from
high- and low-temperature series expansions.
\vfill\eject

\section{Introduction}
In finite-temperature field theory one aims to calculate the effects
of temperature, especially in models that undergo a symmetry
breaking phase transition, such as the standard model.
Direct perturbation theory gives
a series with both ultraviolet (UV) and infrared (IR) divergences.
The former are temperature independent and can be consistently removed
by going from ``bare'' to ``renormalized'' parameters according to
standard renormalization prescriptions.
However, the resulting ``zero-temperature renormalized'' perturbation theory
breaks down in two distinctly different regimes: firstly, when
$T\gg M(T)$, $M(T)$ being a typical finite-temperature mass in the problem;
and secondly when $M(T)\gg M(0)$. The latter can be remedied by expanding
around the finite-temperature mass instead of the zero-temperature mass.
The former occurs in the vicinity of a second, or weakly first order phase
transition, however, the validity of perturbation theory can be restored
by a suitable temperature dependent renormalization.
{}From a physical point of view, the breakdown in perturbation theory occurs
because the effective degrees of freedom in the system near
the phase transition are qualitatively different to those
characteristic of $T\sim0$.
Consequently, perturbative series in terms of the zero-temperature
renormalized parameters provide an inadequate representation
of the temperature-dependent effective degrees of freedom of the system.

Such a breakdown in direct perturbation theory,
with the appearance of IR divergences, is typical of a wide class of
``crossover'' problems wherein the effective degrees of freedom of a system
change radically as a function of scale due to the effect of some
``environmental'' parameter -- in the present case: temperature.
The occurance of such divergences in finite-temperature field theory
has been known since the work of
Dolan and Jackiw \cite{DolanJackiw} and Weinberg \cite{wein}.
The techniques of ``environmentally friendly renormalization''
\cite{NucJphysa,EnvfRG} offer a quite general approach to
investigating crossover behaviour both qualitatively and quantitatively.
In the context of finite-temperature field theory, a temperature dependent
reparametrization defined by normalization conditions provides a method
of tracking the evolution of the effective degrees of freedom
as a function of both scale and temperature.
The resulting finite-temperature renormalization group \cite{FTRG}
is therefore, in principle, environmentally friendly.

In previous work \cite{NucJphysa,EnvfRG,prevfintemp}
a one-parameter family of reparametrizations at
fixed temperature $T$, parameterized by an arbitrary fiducial
finite-temperature mass, was considered. This rendered the complete crossover
for all values of $T$ accessible in one uniform
perturbation expansion. The crossover was analyzed to two loops
and the phase transition shown to be second order, characterized by
three-dimensional critical exponents.
Naturally one would also like to have a reliable description in
terms of the zero-temperature parameters of the theory. Here we
wish to present a complementary approach, where, in the
spirit of the finite-temperature renormalization group, one introduces a
one parameter family of renormalized couplings
parameterized by an arbitrary fiducial temperature $\tau$.
We find that the mass and coupling vanish continuously at $T_c$ and behave
as ${(f_1^{\pm})}^{-1}|T-T_c|^{\nu}$ and $l^{\pm}|T-T_c|^{\nu(4-d_c)}$
respectively,
as expected from critical phenomena (and in agreement with \cite{NucJphysa}).
Here, $d_c$ is the reduced dimension in the critical
regime, $\nu$ and $\eta$ are characteristic exponents,
$f_1^{\pm}$ and $l^{\pm}$ are amplitudes, the $\pm$ refering to above
and below the critical temperature respectively.
In the approximations of this paper, for the physical dimension $d=4$, one
finds
$d_c=3$ and the exponents  take the values
$\nu=1$ and $\eta=0$. Amplitudes differ above
and below the critical point, however certain ratios of these amplitudes,
like critical exponents, are universal numbers.
In the present context this means they are independent of the zero-temperature
mass and coupling which we use to parameterize the theory.
We further use the temperature renormalization
group to calculate the critical temperature as a function of the
zero-temperature parameters of the theory.

The principal new results of the paper are:\\
(i) A prescription for the calculation of the entire crossover,
from zero temperature to the critical regime and beyond, in terms of the zero
temperature parameters and in an expansion free of both
UV and IR divergences. This has been implemented to one loop in the paper.\\
(ii) A perturbatively reliable method for calculation of
the critical temperature.
We find at one loop a critical temperature approximately $20\%$ different
from that obtained by daisy resummation when
the values of the parameters are taken to be reasonable values for the
Higgs sector of the standard model.\\
(iii) The mass, coupling, spontaneous symmetry breaking vacuum
expectation value of the field and the minimum of the effective potential
are obtained as a function of temperature.\\
(iv) Amplitudes in the critical region, ratios of which are in good agreement
with those associated with the three-dimensional Ising model.
The method presented here gives an alternative way of calculating
these quantities to those that exist in the literature to date.
For example we find $f_1^{+}/f_1^{-}=1.92$ which is in good agreement
with the best series expansion results of Liu and Fisher \cite{LiuFisher}
who obtain $1.96\pm 0.01$ (this should be compared with the result $1.41$
obtained from a tree level analysis where fluctuations are ignored
and $1.91$ from the $\epsilon$ expansion at order $\epsilon^2$ \cite{BGZJ}
for $\epsilon=1$).\\
(v) The effective potential and its evolution as a function of temperature,
with emphasis on its convexity.

The structure of the paper is as follows. In Section 2 we present
our finite temperature renormalization prescription. In Section 3
we study the flow equations at one loop and compare our approach with
the work of others. Section 4 is devoted to the critical regime and
the calculation of some amplitude ratios. We study the effective
potential and higher vertex functions in Section 5. The paper ends with
our conclusions and remarks.

\section{Finite-temperature renormalization}
The model we consider is given by the Euclidean action
\be
S[\f_{B}]=\int_0^\b dt\int
d^{d-1}x\left[{1\o2}(\nabla\f_{B})^2+{1\o2}M^2_{B}\f_{B}^2+
{\l_{B}\o4!}\f_{B}^{4}\right]\label{action}
\ee
in terms of bare quantities and the inverse temperature $\b=1/T$.
The effective potential $V$, or effective action
($\G$) per unit volume in constant field, is given
in terms of the renormalized field expectation
value $\fb=\Zf^{-1/2}\fb_B$ in the
presence of some arbitrary constant external current $J$, and obeys
the equation of state
\be
\G^{(1)}=\G^{(2)}_t\fb=J,\label{eqnofstate}
\ee
which specifies $\fb(J)$ and serves to define $\G^{(2)}_t$.
We further define $\G^{(4)}_t$ through
\be
\G^{(2)}=\G^{(2)}_t+{\G^{(4)}_t\o3}\fb^2.\label{dfnofgft}
\ee
Our notation is motivated by considerations of
the $O(N)$ model analytically continued to $N=1$. The structure revealed
is, however, intrinsic to the model under consideration, where more generally
each $n$-point vertex function
admits a decomposition into a polynomial in $\fb$ with coefficients
proportional to $\G^{(n+k)}_t$ where $k$ ranges up to $n$.
We will return to some of these points and
present detailed calculations for the $O(N)$ model elsewhere.

The solution of (\ref{eqnofstate}) on the co-existence curve $J=0$ is $\fm$
and is zero for $T>T_c$, and given by $\G^{(2)}_t=0$ for $T<T_c$.
Expanding around some reference background field $\fb_{H}$, corresponding
to a reference external current $H$, the effective potential can be written as
\be
V(\fb)=U(\fb_{H})+H\Delta\fb
+\sum_{n=2}^\infty{\G^{(n)}\o n!}(\Delta\fb)^n.\label{effpotnl}
\ee
which specifies the renormalized vertex functions $\G^{(n)}$ at zero momentum
as the coefficients in this expansion.
The deviations $\Delta J$ and $\Delta\fb$ from the reference source and
field expectation are related by
\be
\Delta J=\sum_{n=2}^{\infty}{\G^{(n)}\o(n-1)!}(\Delta\fb)^{n-1}\label{delphi}
\ee
Due to the non-analyticity of the effective potential on the co-existence curve
this Taylor expansion strictly only makes sense away from the latter.

Our renormalized parameters are specified by the following
normalization conditions at an arbitrary temperature scale $\tau$, and at
arbitrary $H$
\be
\left.{\p\o\p
p^2}\G^{(2)}_t(p,\fb_H(\tau),M(\tau),\l(\tau),T=\tau)\right|_{p=0}
&=&1,\\
\G^{(2)}(p=0,\fb_H(\tau),M(\tau),\l(\tau),T=\tau)&=&M^2(\tau),\label{massCnd}\\
\G^{(4)}_t(p=0,\fb_H(\tau),M(\tau),\l(\tau),T=\tau)&=&\l(\tau).\label{normcnds}
\ee
Note that the physical mass $m(T)$ at the normalization point
is given by
\be
m^2(\tau)=
{M^2(\tau)\o 1+\left.{\fb_H^2\o3}{\p\o\p p^2}\G^{(4)}_t(\tau)\right|_{p=0}}
\ee
and coincides with $M(\tau)$ when $\fb_H=0$ but not otherwise.

The differential equations which describe an infinitesimal change in
normalization point with fixed bare parameters are
\be
\tau{d\ln Z_\f(\tau)\o d\tau}=\gf,\qquad
\tau{dM^2(\tau)\o d\tau}=\bm,\qquad
\tau{d\l(\tau)\o d\tau}=\bl.
\ee
For $H=0$ the flow functions $\bm$, $\bl$ and $\gf$ take different
functional forms above and below the critical temperature $T_c$,
which is determined by the vanishing of $M$.
The normalization prescription above has the advantage of
preserving, order by order in perturbation theory,
the known symmetry that a given exponent is
independent of whether the critical point is approached
from above or below $T_c$ or the direction in which $H$ is sent to zero.
In deriving them we treat $\l$ in perturbation theory, but
keep $M$ and its derivatives non-perturbatively within this approximation.
More explicitly, in each diagram we eliminate $M_B^2$ in
favour of $M^2$ with the aid of condition (\ref{massCnd}).
This eliminates all diagrams which contain a tadpole
as a sub-diagram.
Subsequently, we differentiate with respect to $\tau$ and solve
for the flow functions, expanding them to the loop order we are working.
We have verified to two-loops that the resulting flow equations are free
of UV and IR divergences.

\section{The Flow Functions and Running Parameters}
We calculate the flow functions to one loop,
taking the reference external current $H=0$, and obtain
\be
\gf & = & 0, \\
\bm & = &\left\{\ba{ll}{\l\o 2} \tau{\p\loop1\o\p\tau},& \tau>T_c\\
& \\
-\l \left(\tau{\p\loop1\o\p\tau}+{3\o2}M^2\tau{\p\loop2\o\p\tau}\right),
&\tau<T_c, \ea\right.\\
\bl & = &-{3\o2}\l^2\tau{d\o d\tau}\loop2 .\label{couplingflow}
\ee
We present these results in diagrammatic form as this renders the structure
of the expressions more readily apparent and easily adaptable
to other situations.
The symbol $\loop{k}$ stands for the one-loop diagram with $k$ propagators,
without vertex factors, at zero external momentum.
It can be obtained from the following basic diagram in $d$ dimensions
\be
\bigcirc&=&
\tau\sum_{n=-\infty}^\infty\int{d^{d-1}k\o(2\pi)^{d-1}}
\ln(k^2+(2\pi n\tau)^2+M^2)\nn\\
&=&-{\G(-{d\o2})M^d\o{(4\pi)}^{d/2}}-{2\tau^d\o{(4\pi)}^{(d-1)/2}\G({d+1\o2})}
\int_0^\infty dq{q^d\o\sqrt{q^2+{z^2}}}{1\o e^{\sqrt{q^2+{z^2}}}-1},
\ee
where $z=M/\tau$, by differentiations with respect to $M^2$.
The first derivative gives
\be
{d\o dM^2}\bigcirc=\loop1,
\ee
whereas for $k\geq1$ we have the general rule that the derivative
with respect to $M^2$ of the loop with $k$ propagators
gives $-k$ times the loop with $k+1$ propagators.

The resulting flow functions above differ crucially with other work.
Compared to the results of Fujimoto et al.
\cite{Fujimoto} for the $\b$-functions, our flow function $\bl$ is
defined with a total derivative rather than a partial one.
The difference between the two flow functions is
\be
\bl-\bl^F=
-{3\o2}\l^2\tau\left({d\o d\tau}-{\p\o\p\tau}\right)\loop2=
-{3\o2}\l^2\tau{dM^2\o d\tau}{\p\loop2\o\p M^2},\label{diffTerm}
\ee
where $\bl^F$ is the flow function as given by Fujimoto et al.
and for small $M/\tau$ takes the form $\bl^F=-{3\l^2\tau\o16\pi M}$.
As the critical temperature is approached both
$\l\hbox{ and }M\sim\tau-T_c$ (see below) which implies
\be
{\bl-\bl^F\o\bl^F}={4\pi\tau\o9M}-{2\o3\pi}+\dots,
\ee
diverges at the critical temperature. The argument is similar in the
broken phase.
The additional term (\ref{diffTerm}) is therefore the dominant contribution
and cannot be neglected.
In contrast, had one dropped this term, one would then find that the coupling
rather than going to zero approaches a constant.
When this solution is inserted into $\bl$ we see that it diverges.
Thus treating $\tau dM^2/d\tau$ as being of higher order
will not be consistent.

A second remark concerns the broken phase.
As is known from previous work \cite{NucJphysa,EnvfRG,prevfintemp,TdrWett}
the coupling $\lambda\rightarrow0$ as one approaches the critical point
from either above or below.
In the present approach this is a direct consequence of the total
derivative in the flow function $\bl$.
In contradistinction, the combined results of
Elmfors and Fujimoto et al.\cite{Fujimoto,Elmfors}
give a discontinuity in the coupling.
Additionally, Elmfors'  results below $T_c$ \cite{Elmfors} differ from ours in
that
his renormalization prescription fails to preserve the aforementioned
symmetry of the critical exponents about the critical temperature.

Of course, the vanishing of $\l$ does not imply the theory becomes
non-interacting.
In dimensional crossover a more natural dimensionless coupling
\cite{NucJphysa,EnvfRG} is the floating coupling $h=4 \l M^2 \loop{3}$.
Using $M$ as a parameter measuring the distance from $T_c$ we obtain
for $h$ the flow equation (for the symmetric phase)
\be
M{\p h\o\p M}=-h+h^2+O\left(\textstyle{M\o T}\right).
\label{uFlow}
\ee
This equation has, in the limit $M\ra0$, a familiar fixed point structure,
with a stable, non-trivial fixed point at $h^*=1$.
Moreover, it has the added advantage of providing a coupling that remains
small for all temperatures, and is proportional to the zero-temperature
coupling in the zero temperature limit.

The differential equation for the coupling
(\ref{couplingflow}) is easy to solve,
since it contains a total derivative, and to the order we are
working takes the same form in both phases.
The solution is
\be
\l^{-1}(\tau)=\l^{-1}(\tau_0)
+{3\o2}\left[\loop2(M(\tau),\tau)-\loop2(M(\tau_0),\tau_0)\right],
\label{lambdaSolution}\ee
This expression is now manifestly finite in four dimensions, where
$\loop2$ diverges logarithmically.
To ensure that the same initial conditions are imposed on both sides of $T_c$
one may use the requirement that the bare coupling $\l_B$ is the same in both
phases.
Eliminating $\l_B$ from the two dimensionally regulated expressions,
$\l^{-1}_\pm(\tau)=\l^{-1}_B+{3\o2}\loop2(M_\pm(\tau),\tau)$,
then gives $\l_+$ in terms of $\l_-$. To one loop we find
\be
\l^{-1}_+(\tau_+)=\l^{-1}_-(\tau_-)+{3\o2}
\left[\loop2(M_+,\tau_+)-\loop2(M_-,\tau_-)\right].
\ee
This solution may now be substituted into the differential equation
for $M(\tau)$, which we solve numerically.

After solving the flow equations we are free to choose the reference
temperature $\tau$ equal to the actual temperature $T$ of interest.
In fact this is essential if one wishes to obtain perturbatively sensible
results for physical quantities \cite{Filipe}.
Because of the renormalization conditions (\ref{massCnd},\ref{normcnds})
the parameters $M(T)$ and $\lambda(T)$ therefore describe the behaviour
of the vertex functions $\G^{(2)}$ and $\G^{(4)}_t$ at zero momentum.
We present the numerical integration of the differential equations
in the figures.
Figure 1 shows the behaviour of $m(T)$ and $\l(T)$,
from zero temperature in the broken phase up to
temperatures above the critical temperature.
With these equations one is also able to determine the critical temperature
in terms of the zero-temperature parameters $m(0)$ and $\l(0)$.
As may be expected on dimensional grounds $T_c$ is proportional to $m(0)$,
the constant of proportionality being a function of $\l(0)$.
To facilitate comparison with the literature we have plotted
$t:=T_c\sqrt{\l(0)}/M(0)$ versus $\l(0)$ in Figure 2.
The original result of Dolan and Jackiw \cite{DolanJackiw} is then
a horizontal line at $t=\sqrt{12}$ in this graph,
where their renormalized parameters are interpreted
as our zero-temperature parameters $M(0)$ and $\l(0)$.
Our critical temperature is larger than theirs
and the value $t=\sqrt{12}$ is approached in the zero-coupling limit.
Amelino-Camelia and Pi \cite{AmelinoCameliaPi}, who found
the transition to be first order, have a value of
$t=4.901$ at $\l(0)=0.05$, which is even larger than
our equivalent value $t=3.607$.
We have also plotted the explicit result given by Lawrie \cite{Lawrie}.
His curve for $t$ starts off in the negative direction,
irrespective of the linear term of which the prefactor was not determined.
Additionally, we compare with ``coarse-graining''
type renormalization groups such as used in \cite{TdrWett}. There the critical
temperature is found to increase as a function of $\lambda(0)$ but at a
rate substantially less than we find. Also their results are dependent on
the particular type of cutoff function used.

If we take values for $\l(0)$ and $M(0)$ to be those associated with
estimates for the equivalent parameters in the Higgs sector of the
standard model \cite{FordJonesetc} we find, for $\l(0)=1.98$ and $M(0)=200$GeV
that $T_c=613$GeV. By comparison Dolan and Jackiw's result gives $492.4$GeV.
For $\l(0)=3.00$ and $M(0)=246$GeV
the corresponding results are $T_c=639$GeV and $492.0$GeV respectively.

Knowing the behaviour of $M(T)$ and $\l(T)$ we are able to plot
the field expectation value
\be
\fm=\left\{\ba{ll}0,&T\geq T_c\\
\sqrt{{3\o\l}}M(1+\dots),&
T<T_c\ea\right.
\ee
versus $T$, see Figure 3.
At the critical temperature $\fm$ is seen to vanish continuously, once again
confirming that the phase transition is second order.

\section{Critical behaviour}
Since $M(T)\ll T$ near $T_c$ the finite-temperature four-dimensional theory
should reduce there to a three-dimensional Landau-Ginzburg model,
in accord with finite size scaling (see \cite{Barber} for a review),
with vertex functions behaving as
\be
\G^{(n)}_{\pm}=\gamma^{(n)}_{\pm}|T-T_c|^{\nu\left(d_c-n{d_c-2+\eta\o2}\right)}.
\label{scaling}
\ee
where $d_c=d-1$ is the reduced dimension at the critical point, $\nu$ and
$\eta$ are
critical exponents and $\gamma^{(n)}_{\pm}$ are amplitudes.

We now investigate more explicitly the critical regime.
We use the expansion formula
\be
\bigcirc=T^d\left[-{2\G(d/2)\zeta(d)\o\pi^{d/2}}
+{\G({d-2\o2})\zeta(d-2)\o 2\pi^{d/2}}{z^2}
-{\G({1-d\o2})\o {(4\pi)}^{(d-1)/2}}z^{d-1}\right.\nn\\
\left.-{\G({d-4\o2})\zeta(d-4)\o{(4\pi)}^2\pi^{(d-4)/2}}z^4
-2\pi^{d/2}{\displaystyle\sum_{m=1}^\infty}
{(-1)}^m{\G(m+{5-d\o2})\zeta(2m+5-d)\o\G({1\o2})\G(m+3)}
{\left({z\o2\pi}\right)}^{2m+4}\right]
\ee
for any dimension $d$ between 3 and 4.
As noted by Dolan and Jackiw \cite{DolanJackiw}, for $d=4$
the tadpole $\loop1$ has in this limit a quadratic $T$-dependence
\be
T{\p\loop1\o\p T}={T^2\o6}-{MT\o4\pi}+\dots,
\label{highTtadpole}
\ee
while all one loop diagrams with more propagators grow only linearly
in $T$ according to
\be
T{\p\loop{k}\o\p T}={\G(k-3/2)\o(4\pi)^{3/2}(k-1)!}\:{T\o M^{2k-3}}+
\dots,\qquad k\geq2.
\label{highTloop}
\ee
In the neighbourhood of $T_c$ for the four-dimensional theory,
we find therefore for the flow functions
\be
\gf&=&0,\\
\bm&=&\left\{\ba{ll}\l\left({T^2\o12}-{MT\o8\pi}+\dots\right),&T\downarrow
T_c\\
&\\
-\l\left({T^2\o6}-{29MT\o64\pi}+\dots\right),&T\uparrow T_c\ea\right.\\
\bl&=&{3T\l^2\o16\pi M}\left({\l T^2\o24M^2}-1+\dots\right),
\ee

In Figure 1 it is seen that, as the critical point is approached,
the coupling vanishes in accord with previous results
\cite{NucJphysa,EnvfRG,prevfintemp,TdrWett}.
More specifically this also follows from the solutions (\ref{lambdaSolution})
by noting that the function $\loop2$ behaves like $T/8\pi M$.
Hence, near the critical temperature
\be
\l_+={16\pi M_+\o3T}+\dots,\qquad M_+={2\pi\o9}(T-T_c),\qquad T>T_c;\nn\\
\l_-={16\pi M_-\o3T}+\dots,\qquad M_{-}={4\pi\o 9}(T_c-T),\qquad T<T_c.
\label{ampRatio}\ee
These take the form $M^2_{\pm}={(C^{\pm})}^{-1}|T-T_c|^{\gamma}$ with
$\l_{\pm}=l^{\pm}|T-T_c|^{\nu}$ and $m_{\pm}={(f_1^{\pm})}^{-1}|T-T_c|^{\nu}$,
where we use the notation of Liu and Fisher \cite{LiuFisher} for the
amplitudes.
As $\gamma=\nu(2-\eta)$ we see that $\nu=1$ and $\eta=0$,
and that the amplitude ratios at two points symmetric
around $T_c$ are
\be
{C^+\o C^-}=4,\qquad\qquad
{f_1^+\o f_1^-}=2\sqrt{12\o13}\approx1.92,\qquad\qquad
{l^+\o l^-}={1\o2},
\ee
which is indicative of a cusp in the mass and coupling as the theory passes
through the critical temperature, as can be seen in Figure 1.
These ratios are universal numbers analogous to the critical exponents.
The exponents $\nu$ and $\eta$ have been estimated by different methods
(see \cite{ZinnJustin}) with the results $\nu=0.6310\pm 0.0015$
and  $\eta=0.0375\pm 0.0025$.
The best estimates for the amplitude ratios are the high- and low-temperature
series expansion results of Liu and Fisher \cite{LiuFisher} who find
\be
{C^+\o C^-}=4.95\pm0.15,\qquad\qquad
{f_1^+\o f_1^-}=1.96\pm0.01,\qquad\qquad
\ee
which our results are in good agreement with. By comparison:
at tree level  (mean field theory) $C^+/C^-=2$ and $f_1^+/f_1^-=1.41$,
whilst in the $\epsilon$ expansion at order $\epsilon^2$,
assuming dimensional reduction,
$C^+/C^-=4.8$ and $f_1^+/f_1^-=1.91$ \cite{BGZJ}.

We see here that the amplitude ratios are substantially better than the
results for critical exponents. This underlines the notion of complimentarity
between the current approach of flowing the environment and that of
\cite{NucJphysa,EnvfRG} where the flow parameter was the finite
temperature mass.  At one loop the latter group gives better results for
exponents whereas the former gives better results
for exponents. One would expect that at higher loop orders, with suitable
resummations, the different schemes will converge to the same results.

\section{Effective potential}
Let us now construct the effective potential (\ref{effpotnl})
at the normalization point $T=\tau$.
We determine the minimum of the effective potential $U$, in the absence
of external currents $H=0$, by using the flow equation
\be
\tau{dU(\tau)\o d\tau}=\b_U={1\o2}\tau{\p\bigcirc\o\p \tau}+\dots.
\ee
Figure 4 shows the minimum $U(T)$ of the effective potential relative
to the minimum at zero temperature $U(0)$.
In the setting of the early universe, normalizing the latter to zero would
correspond to a vanishing cosmological constant.
In the high-temperature limit, $T$ is much larger than
both $M(0)$ and $T_c$,
the effective potential $U$ is relatively close to that of an ideal-gas
$-\pi^2T^4/90$ \cite{DolanJackiw}.
Since the mass of the effective excitations increases
approximately linearly with $T$
further increase of temperature will not give rise
to a regime where dimensional reduction is valid,
rather what will happen is that quantum corrections will be
suppressed due to a large mass in the diagrams.
However, the effective potential should still have the form $\sigma T^4$
with some non-ideal Stefan-Boltzmann constant $\sigma$.

The first few of the remaining vertex functions involved are
\be
\G^{(3)}=\left\{\ba{ll}0,&T>T_c\\
\sqrt{3\l}M\left(1+3\l M^2\loop{3}\right),
&T<T_c,\ea\right.\\
\G^{(4)}=\left\{\ba{ll}\l,&T>T_c\\
\l\left(1+18\l M^2\loop{3}-27\l M^4\loop{4}\right),&T<T_c,
\ea\right.\\
\G^{(5)}=\left\{\ba{ll}0,&T>T_c\\
3\sqrt{3}{\l}^{5/2}M\left(5\loop3-30M^2\loop4+36M^4\loop5\right),&T<T_c,
\ea\right.\\
\G^{(6)}=\left\{\ba{ll}15\l^3\loop3,&T>T_c\\
15\l^3\left(\loop3-27M^2\loop4+108M^4\loop5-108M^6\loop6\right),&T<T_c,
\ea\right.
\ee
and are plotted in Figure 5.
We see from (\ref{ampRatio}) that near the critical temperature
$\G^{(3)}\sim |T-T_c|^{3/2}$ and $\G^{(5)}\sim|T-T_c|^{1/2}$ for $T>T_c$, but
are zero for $T<T_c$, whilst $\G^{(4)}\sim|T-T_c|$
in accordance with (\ref{scaling}).
The vertex function $\G^{(6)}$ is the first one that remains non-zero
at the critical temperature, however, this is an artifact of the fact that at
this order $\eta=0$. At higher orders all vertex functions $\G^{(n)}$
for $n\geq6$ will in fact diverge as the critical temperature is approached.

As one approaches the critical temperature we find to this order
\be
\G^{(4)}_{+}={16\pi M_{+}\o3 T},
\qquad\G^{(4)}_{-}={28\pi M_{-}\o3T}
\qquad\hbox{with}\quad{\gamma^{(4)}_{+}\o\gamma^{(4)}_{-}} ={2\o7},\\
\G^{(6)}_{+}=T_c^{-2}{640\pi^2\o9},
\qquad\G^{(6)}_{-} =-T_c^{-2}{1520\pi^2\o9}
\qquad\hbox{with}\qquad{\gamma^{(6)}_+\o\gamma^{(6)}_{-}}=-{8\o19}
\ee
In the broken phase the six-point function turns out to be negative, but one
should remember that there are an infinite number of vertex functions
and a truncation of the effective potential at a low
order is only appropriate for sufficiently small $\Delta\fb$
and for temperatures sufficiently far from the critical temperature.
One can see in Figure 6 that the truncation at $\G^{(6)}$ breaks down
for smaller values of $\Delta\fb$ the closer $T_c$ is approached.

The complete shape of the effective potential can be thought
of as that of a boat where the lower hull has been sawn off.
The flat base of the boat has a side projection whose characteristic shape
divided by $T^4$ is shown in Figure 4. The flat region slopes downwards as
the temperature is increased and ends at the critical temperature.
A top projection of the bottom of the boat can be seen from Figure 3
when the curve $\fb_0$ is thought of as being reflected around the $T$ axis.
One sees that as a positive reference external current $H$ is sent to zero for
$T<T_c$ the minimum of the effective potential occurs at
a non-zero positive value of the field expectation $\fb_0(T)$.
If the field were reversed in sign and sent to zero
the field expectation would be $-\fb_0(T)$, the width of the boat
across the base being $2\fb_0(T)$. The bottom of the boat represents
the co-existence region of the phase diagram and $\fb_0(T)$ should
be connected with $-\fb_0(T)$ for the same value of $T$ by a ``tie-line''.
The critical point is the front of the boat and is sewn on in a highly
non-analytic way, as indeed is the entire base.
We retain all these features in our approach.

\section{Conclusions}
In this paper we have considered an environmentally friendly renormalization
group complementary to that considered in
\cite{NucJphysa,EnvfRG,prevfintemp}. Here, it is the
environment itself, i.e. the temperature, which defines a one parameter
family of reparametrizations. By running the environment we can answer
questions relatively simply that were difficult to answer whilst running
the finite-temperature mass. The chief example of which, considered
here, is the determination of the critical temperature. The formalism
we have exhibited is quite capable of calculating the latter to
any order in perturbation theory. Preliminary two loop results have been
obtained and will be presented elsewhere. We emphasize
that we determine the critical temperature and critical amplitudes,
which are intrinsically ``non-universal'' concepts,
using only the renormalization group. The critical temperature is
determined by the integrals of the characteristic equations and is
sensitive to the non-universal initial conditions of these equations.
Universal quantities in contrast are determined by the characteristic
functions at the critical temperature, and as the name suggests
are insensitive to initial conditions.

The formalism herein, also
illuminates a crucial difference between environmentally friendly
renormalization and coarse-graining procedures. In the latter
only the coarse-graining scale defines a renormalization group and one
is restricted to one flow parameter.
In environmentally friendly renormalization one can, in principle, have
a multi-parameter flow
with as many parameters as the dimension of the space of couplings.
In contrast to the coarse-graining approach, here only one single initial
condition need be given for the coupling constants. We believe that
this is a significant advantage of reparametrization over coarse graining.

Finally, to investigate the full standard
model as opposed to just the Higgs sector one needs to confront the finite
temperature behaviour of the non-Abelian gauge fields. As discussed in
\cite{qcd}, for temperatures much larger than other scales the gauge coupling
in the magnetic sector grows very large. This is a signal that collective
degrees of freedom other than quarks and gluons are playing a significant
role. To address this issue one needs to develop an environmentally friendly
renormalization that is capable of tracking the change between quark-gluon
degrees of freedom and hadron-meson degrees of freedom. We hope to be able to
return to this very interesting question in the future.

\section*{Acknowledgements}
C.R.\ Stephens has been supported by an EU human capital and mobility
fellowship. Denjoe O'Connor has received financial support from NWO and
expresses his gratitude to the Inst.\ for Theor.\ Physics,
University of Amsterdam where part of this work was done, for its hospitality
and wishes to thank P.J.\ Upton for bringing \cite{LiuFisher} to his attention
and C.\ Ford for a discussion on the bounds for the Higgs parameters.
M.A.\ van Eijck wishes to thank I.D.\ Lawrie for illuminating discussions
and P.\ Elmfors for his help in clarifying issues associated with his coupling
in the two phases.

\newpage
\section*{Figure Captions}
\begin{description}
\item{Fig.\ 1:}
The mass $m(T)$ and coupling $\l(T)$ as a function of the temperature $T$.
The graphs are obtained by solving the flow functions with initial
(zero-temperature) coupling $\l(0)=1$, starting in the broken phase.
$T$ and $m(T)$ are given in units of $m(0)$.
The critical temperature $T_c=4.080 m(0)$ separates the two phases.
\item{Fig.\ 2:} The critical temperature $T_c$ as a function of the
zero-temperature parameters $\l(0)$ and $M(0)$.
$T_c$ is proportional to $M(0)$.
Graph (a) is obtained from our numerical solution,
(b) is the early result of Dolan and Jackiw \cite{DolanJackiw},
and (c) shows the expression obtained by Lawrie \cite{Lawrie}
(up to a linear term).
The crosses and stars represent values obtained
by Tetradis and Wetterich \cite{TdrWett} for two specific choices
of coarse-graining, corresponding to their tables 2 and 6 respectively.
\item{Fig.\ 3:} The field expectation value $\fb$ as function of
the temperature $T$ corresponding to the solution of Fig.\ 1.
\item{Fig.\ 4:} The minimum of the effective potential $U$
as function of the temperature $T$,
corresponding to the solution of Fig.\ 1,
normalized by the ideal gas value.
\item{Fig.\ 5:} The vertex functions $\G^{(2)},\dots,\G^{(6)}$
to one-loop at zero momentum, corresponding to the solutions of Fig.\ 1.
\item{Fig.\ 6:} The source $\Delta J$ as a function of $\Delta\fb$
for the temperatures $T/m(0)=0,3,4,6$,
corresponding to the solution of Fig.\ 1 and truncated beyond the
contribution from $\G^{(6)}$. The critical temperature is at $T_c=4.080 m(0)$.
\end{description}

\end{document}